\documentclass[12pt]{article}
\usepackage{amsfonts,amsmath}
\usepackage{hyperref}

\makeatletter \@addtoreset{equation}{section} \makeatother

\newcommand{\be}{\begin{equation}}
\newcommand{\ee}{\end{equation}}
\newcommand{\bee}{\begin{eqnarray}}
\newcommand{\beee}{\begin{array}}
\newcommand{\eee}{\end{eqnarray}}
\newcommand{\eeee}{\end{array}}

\newcommand{\un}{{\underline{n}}}
\newcommand{\um}{{\underline{m}}}

\newcommand{\ga}{\alpha}
\newcommand{\pa}{{\dot{\ga}}}
\newcommand{\pb}{{\dot{\gb}}}

\newcommand{\gb}{\beta}

\newcommand{\ie}{{\it i.e.,} }

\newcommand{\go}{\omega}

\newcommand{\q}{\,,\qquad}

\newcommand{\nn}{\nonumber}

\newcommand{\half}{\frac{1}{2}}

\newcommand{\p}{\partial}

\newcommand{\Y}{{\mathcal{Y}}}

\def\l{\lambda}

\newcommand{\dga}{{\dot \ga}}

\newcommand{\dr}{{\rm d}}

\begin{document}

\begin{flushright}

{\small FIAN/TD/07-19}
\end{flushright}
\vspace{1.7 cm}

\begin{center}
{\large\bf Field Equations for the Simplest Multi-Particle Higher-Spin Systems}

\vspace{1 cm}

{\bf  I.S.~Degtev and M.A.~Vasiliev}\\
\vspace{0.5 cm}
{\it
 I.E. Tamm Department of Theoretical Physics, Lebedev Physical Institute,\\
Leninsky prospect 53, 119991, Moscow, Russia}

\end{center}

\vspace{0.4 cm}

\begin{abstract}
\noindent
We derive the product law for the simplest multi-particle higher-spin algebra $M_2(A)$
        and its factor-algebra $M^2(A)$ with rank-one fields factored out.
        Equations of motion for the
         systems resulting from these algebras are analysed.
        We conclude that the equations for $M_2(A)$ describe the conformal off-shell higher-spin
         system with the rank-two
        fields representing the off-shell degrees of freedom of the originally massless system. On the other hand
        the equations resulting from $M^2(A)$ describe the infinite system of conserved currents unrelated to massless fields.

\end{abstract}

\newpage
\tableofcontents

\newpage

\section{Introduction}

One of the major    problems of the theory of fundamental interactions
is to understand relation
between String Theory \cite{Green:1987sp}, that contains infinite towers of massive higher-spin (HS) fields, and HS gauge theories in which all HS fields are massless and which exhibit infinite-dimensional
HS symmetries (for  review see, e.g., \cite{Vasiliev:1999}). Arguments that String Theory
possesses higher symmetries in the high-energy
limit were given long ago in \cite{Gross:1987ar,Gross:1988ue}. This suggests that String Theory
should be related to HS theory. For related discussion see, e.g.,
\cite{Metsaev:1999kb,Sagnotti:2013bha}.

More recently, further proposals  on the relation
between HS gauge theories and String Theory were put forward in
\cite{Gaberdiel:2015wpo,Gaberdiel:2018rqv}.
 In \cite{Vasiliev:2018zer} it was conjectured
 that there exists a broad class of HS  gauge theories
based on the HS algebras and their further multi-particle extensions.
In \cite{Vasiliev:2018zer} it was argued that dynamics based on certain algebras
has no room for usual massless fields while other models properly describe HS gauge fields.
 In this paper, we consider in
some detail  how this can  be seen at the level of field equations associated with
 one or another algebra in the framework of the simplest multi-particle extension  proposed
 originally in \cite{Vasiliev:2012tv}.

Specifically, we consider two versions of the two-particle algebra and discuss the form of the
related field equations. The corresponding  algebras denoted $M_2$ and $M^2$, respectively,
are  factor-algebras of the
simplest multi-particle algebra of \cite{Vasiliev:2012tv}.
To these algebras we associate two HS systems and derive their equations of motions.
It will be shown that the $M_2$ system properly describes massless HS fields as well
as two-particle fields that can be interpreted as conserved currents.
In particular, it will be explained how the $M_2$ system is related to the system of
HS currents built from products of HS fields according to \cite{Gelfond:2003vh,Gelfond:Current}.
On the other hand, it will be explicitly shown that,
 in agreement with the general group-theoretical argument of \cite{Vasiliev:2018zer},
  the $M^2$ system, that can be
understood as resulting from the
oscillator (Weyl) algebra with two sets of oscillators, cannot describe massless
fields.

The rest of the paper is organized as follows. In Section \ref{M}  the construction
of the multi-particle algebras is recalled. In Sections \ref{M2d} and \ref{M2u} we analyse
field equations of the $M_2$ and $M^2$ systems, respectively. Brief conclusions are in
Section \ref{con}. Appendix contains some details of the derivation of the
relevant multi-particle algebras.

\section{Multi-particle algebras}
\label{M}

First, we recall relevant elements of the algebraic construction of
\cite{Vasiliev:2012tv}. Let $A$ be some associative
algebra $A$ with basis elements $t_i$, the product law $\star$  and  structure
coefficients
\be
    t_i \star t_j = f_{i j}^k t_k
\ee
obeying associativity condition
\be
    (t_i \star t_j) \star t_k = t_i \star (t_j \star t_k) \in A, \quad t_i, t_j, t_k \in A\,.
\ee
 $A$ is also assumed to be
unital with the unit element $e_{\star}$ obeying
\be
    e_{\star} \star t = t \star e_{\star} = t, \quad \forall t \in A\,.
\ee

For any such $A$  it is possible to build an associative multi-particle algebra $ M(A)$
having the meaning of the universal enveloping of $A$ (more precisely, of
the Lie algebra associated with $A$). As a linear space
\be
    M(A) = \bigoplus_{n=0}^\infty Sym \,\underbrace{A\otimes ... \otimes A}_{n}\,,
\ee
{\it i.e.}, the basis for this algebra consists of symmetric monomials.\footnote{For simplicity,
in this paper we consider purely bosonic models. To include fermions one has to consider
an appropriate graded extension of the multi-particle algebra introduced in \cite{Vasiliev:2018zer}.}
\be
    T_{i_1, \dots, i_n} = Sym\, t_{i_1} \otimes\dots \otimes t_{i_n}\,, \quad T_{\dots j\dots k\dots} = T_{\dots k\dots j\dots}\,.
\ee
The monomial of  degree zero in $t_i$ is identified with the unity $Id$ of $M(A)$.

Due to symmetrization, elements of  $M(A)$ can be represented as functions of the
commuting variables $\alpha_i$
\be
\label{F}
    F = \sum_{n=0}^\infty F^{i_1, \dots, i_n}T_{i_1 \dots, i_n} \sim F(\alpha) = \sum_{n=0}^\infty F^{i_1, \dots, i_n} \alpha_{i_1}\dots\alpha_{i_n}\,.
\ee

Associative product $\circ$ in $M(A)$ is generated by the $\star$ product in $A$ as follows
\cite{Vasiliev:2012tv}
\be\label{prod}
    F(\alpha)\circ G(\alpha) = F(\alpha)\exp\left(\frac{\overleftarrow{\partial}}
    {\partial \alpha_i}f_{ij}^n\alpha_n\frac{\overrightarrow{\partial}}{\partial \alpha_j}
    \right)G(\alpha)\,,
\ee
where $f_{ij}^n$ are the structure constants of  $A$ and derivatives $\frac{\overleftarrow{\partial}}{\partial \alpha_i}$ and $\frac{\overrightarrow{\partial}}{\partial \alpha_j}$ act on $F(\alpha)$ and $G(\alpha)$, respectively.

One can easily check that  associativity of the product $\circ$ in  $M(A)$ follows from that of $A$
\be
    (F_1(\alpha) \circ F_2(\alpha)) \circ F_3(\alpha) = F_1(\alpha)\circ (F_2(\alpha) \circ F_3(\alpha))\,.
\ee

$M(A)$  has a family of two-sided ideals $\mathcal{I}^M$:
\be
     \{F\in  \mathcal{I}^M:
 F(\alpha) = \sum_{n=M}^\infty F^{i_1, \dots, i_n} \alpha_{i_1}\dots\alpha_{i_n}\}\,.
\ee
Specifically, in this paper we are interested in the ideal $\mathcal{I}^3$  spanned
by polynomials of degree three or higher.

The factor-algebra $M_2(A):= M(A)/\mathcal{I}^3$ consists of polynomials of degrees
$p\leq 2$ with the basis monomials $1$, $\alpha_i$, $\alpha_i\alpha_j$.
Discarding polynomials of degree three or higher, the product law (\ref{prod}) induces
the following product law $\overset{M_2}{\circ}$ in $M_2$
\be
    \alpha_{i_1} \overset{M_2}{\circ} 1 = \alpha_{i_1} \exp\left(\overleftarrow{\partial_m}f_{ml}^k \alpha_k \overrightarrow{\partial_l}\right) 1 = \alpha_{i_1}\,,
\ee
\be
        \alpha_{i_1} \overset{M_2}{\circ} \alpha_{i_2} = \alpha_{i_1}  \alpha_{i_2} + \alpha_{i_1} \star \alpha_{i_2}\,,
\ee
\be
        \alpha_{i_1}\overset{M_2}{\circ} (\alpha_{i_2} \alpha_{i_3}) = (\alpha_{i_1} \star \alpha_{i_2}) \alpha_{i_3} + (\alpha_{i_1} \star \alpha_{i_3}) \alpha_{i_2}\,,
\ee
\be
    (\alpha_{i_1} \alpha_{i_2}) \overset{M_2}{\circ} \alpha_{i_3} = \alpha_{i_1} (\alpha_{i_2} \star \alpha_{i_3}) + \alpha_{i_2} (\alpha_{i_1} \star \alpha_{i_3})\,,
\ee
\be
        (\alpha_{i_1} \alpha_{i_2}) \overset{M_2}{\circ} (\alpha_{i_3} \alpha_{i_4}) = (\alpha_{i_1}\star\alpha_{i_3})(\alpha_{i_2}\star \alpha_{i_4}) + (\alpha_{i_1}\star \alpha_{i_4})(\alpha_{i_2}\star\alpha_{i_3})\,.
\ee

Algebra $M_2$ contains unit element $Id$ (which is $1$ in our notation) but also contains the element $e_{\star}$ which was unit in the algebra $A$ with respect to
 the product $\star$. These two elements generate another two-sided
 ideal:
\be
    \mathcal{I}_{e_{\star} - 2Id}: \{\alpha: \alpha = (e_{\star} - 2 Id)
    \overset{M_2}{\circ} G, \quad \forall G\in M_2(A)\}\,.
\ee
The coefficient $2$ is prescribed because, as shown in \cite{Vasiliev:2012tv}, otherwise it would not be an ideal. For example,
\be
    e_{\star} \overset{M_2}{\circ} f^2 = e_{\star}f^2 + 2(e_{\star}\star f)f \sim 2f^2
\ee
since $e_{\star}f^2 \in \mathcal{I}_{e_{\star} - 2Id}$.

This ideal gives rise to another factor-algebra $M^2(A) := M_2(A) /
\mathcal{I}_{e_{\star} -  2Id}$. We can chose the following  basis in the ideal:
\be
    \alpha_0' \equiv e_{\star} - 2 Id \in \mathcal{I}_{e_{\star} - 2Id}\,,
\ee
\be
    \alpha_i' \equiv (e_{\star} - 2Id) \overset{M_2}{\circ} \alpha_i = e_{\star}\alpha_i - \alpha_i \in \mathcal{I}_{e_{\star} - 2Id}\,,
\ee
\be
    (e_{\star} - 2 Id)\overset{M_2}{\circ}\alpha_i\alpha_j = 0 \in \mathcal{I}_{e_{\star} - 2Id}\,.
\ee
The algebra $M_2(A)$ spanned by $1$, $\alpha_i$, $\alpha_i\alpha_j$
 has dimension  $\half n(n+1) + n + 1$ for $i,j=1,\dots, n$.
The ideal has  dimension  $n + 1$. Its basis can be
completed to that of $M_2(A)$  using elements $\alpha_i\alpha_j$. Thus,
$M^2(A) = M_2(A) / \mathcal{I}_{e_{\star} - 2Id}$ can be represented by
  the elements $\alpha_i\alpha_j$ with the  product law
\be
    (\alpha_{i_1} \alpha_{i_2}) \circ (\alpha_{i_3} \alpha_{i_4}) = (\alpha_{i_1}
    \star\alpha_{i_3})(\alpha_{i_2}\star \alpha_{i_4}) + (\alpha_{i_1}\star \alpha_{i_4})(\alpha_{i_2}\star\alpha_{i_3})\,,
\ee
which is the product law of the oscillator (\ie Weyl) algebra with the doubled set of
oscillators.

Now we are in a position to write down the field equations for the HS systems associated with
$M_2(A)$ and $M^2(A)$.

\section{$M_2$  system  }
\label{M2d}
\subsection{Higher-spin algebra}
\label{HSA}

The consideration of the previous section was applicable to any
associative  algebra $A$. Now we identify $A$ with the simplest
 HS algebra (see e.g. \cite{Vasiliev:1999} and references therein), which is
 the associative algebra of polynomials  $F(Y;K) \in A$
\be
    F(Y;K) = \sum_{n=0}^\infty\sum_{p,q=0,1} F_{pq}^{A_1\dots A_n} Y_{A_1}\cdot \dots \cdot Y_{A_n} k^p\bar k^q,
    \quad A_1, \dots, A_n=1,\dots, 4
\ee
with the star product of $K=(k,\bar k)$--independent functions
\be
\label{star}
    (F_1 \star F_2) (Y) = F_1(Y) \exp[i\overleftarrow{\partial}^{A}C_{AB}\overrightarrow{\partial}^{B}]F_2(Y)\,,
\ee
where ${\partial}^{A}:= \frac{\partial}{\partial Y_A}$ and
\be
    C_{AB} = -C_{B A}\,
\ee
is a non-degenerate symplectic form. In other words, the algebra of $K$--independent
functions  is the Weyl algebra $A_2$. The presence of the Klein operators
$K=(k,\bar k)$, which extends $A_2$ to its semidirect product with the group algebra of
$\mathbb{Z}_2\times \mathbb{Z}_2$, is important in many respects and, in particular,
for the formulation of nonlinear HS equations of \cite{Vasiliev:1988sa,Vasiliev:1992av}. The product law in the resulting algebra $A$
supplements (\ref{star}) by the relations
\be
\label{K}
k\star F(y,\bar y) = F(-y,\bar y)\star k\q \bar k\star F(y,\bar y) =
F(y,-\bar y)\star \bar k\q k\star k= \bar k\star \bar k = 1\,,\quad k\star \bar k =
\bar k\star k\,,
\ee
where $y_\ga$ and $\bar y_\dga$ are the left and right spinor components of
$Y_A=(y_\ga\,,\bar y_\dga)$, $\ga,\dga = 1,2$.
In other words $k$ and $\bar k$ generate automorphisms of the Weyl algebra $A_2$,
that change signs of the left and right spinors, respectively.

Denoting  $Y_A$, $K$ by  $\mathcal{Y}= (Y_A, K)$, algebra $A$
is the algebra of functions $f(\mathcal{Y})$. As a linear space $A$ can be
decomposed into direct sum of its even and odd parts
\be
A=A^0 \oplus A^1
\ee
formed, respectively, by the elements $f(\Y)$ even and odd under the
automorphism
\be
\tau f(Y;k,\bar k) = f(Y;-k,-\bar k)\,,
\ee
\ie
\be
f_0 (Y;k,\bar k)\in A^0:\qquad f_0(Y;k,\bar k) =f_0(Y;-k,-\bar k)\,,
\ee
\be
f_1 (Y;k,\bar k)\in A^1:\qquad f_1(Y;k,\bar k) =-f_1(Y;-k,-\bar k)\,.
\ee

Clearly, $A^0$ forms a subalgebra of $A$ which we call proper HS algebra.
The one-form gauge fields of this algebra
\be
\label{1hs}
\go_{hs}(Y;K|x)=\go_{hs}(Y;-K|x)
\ee
describe the genuine massless
HS fields of all spins $s\geq 1$ including the spin-two field describing
the graviton \cite{Vasiliev:1988sa}.
The gauge fields
\be\go_{top}(Y;K|x)=-\go_{top}(Y;-K|x)
\ee
 associated with $A^1$ describe an infinite set of
topological gauge fields, each describing at most a finite number of
degrees of freedom. (For details see \cite{Vasiliev:1988sa,Vasiliev:1999}).

Another fundamental object of the HS theory
is a zero-form $C(Y;K|x)$ related to $\go(Y;K|x)$ by virtue of the field equations
of \cite{Vasiliev:1988sa}. In this case however, massless (topological)  fields are
associated with
odd (even) zero-forms   $C(Y;K|x)$
\be
\label{0hs}
C_{hs}(Y;K|x)=-C_{hs}(Y;-K|x)\,,
\ee
\be C_{top}(Y;K|x)=C_{top}(Y;-K|x)\,.
\ee
Note that zero-forms $C_{hs}(Y;K|x)$ belong to so-called twisted
adjoint module over the proper HS algebra. In this paper we focus on
the equations on the HS fields neglecting possible contributions to
the sector of topological fields.

\subsection{$M_2$}
Then  elements of $M_2(A)$ can be represented by the unit element $Id$ and
symmetric polynomials of two types%
\be
    F(\Y)
\ee
and
\be
    F(\Y^1; \Y^2) = F(\Y^2; \Y^1)\,.
\ee
The product law is
\be
\label{circ_a}
    F(\Y) \circ G(\Y) = F(\Y^1) G(\Y^2) + F(\Y^2)G(\Y^1) + F(\Y) \star G(\Y)\,,
\ee
\be
    F(\Y) \circ G(\Y^1; \Y^2) = F(\Y^1) \star_{1, 1} G(\Y^1; \Y^2) + F(\Y^2)\star_{2, 2}G(\Y^1; \Y^2)\,,
\ee
\be
\label{circ_b}
    F(\Y^1; \Y^2)\circ G(\Y^1; \Y^2) = F(\Y^1; \Y^2)\star_{1,1} \star_{2, 2} G(\Y^1; \Y^2)\,,
\ee
where star product for the $Y$ variables is defined as
\be
    \star_{i, i} = \exp\left[i
    \frac{\overleftarrow{\partial}}{\partial Y^{i}_{A}}C_{AB}
    \frac{\overrightarrow{\partial}}{\partial Y^{i}_{B}}\right], \quad i={1, 2}\,.
\ee
(No summation over $i$. For more detail on the derivation of this product law see Appendix.)
Analogously,
\be
K_i \circ \Y^j = \Y^j \circ K_i \q i\neq j\,,
\ee
while the product law of $K_i$ with $\Y_i$ at the same $i$ keeps the form (\ref{K})
unchanged.

With this associative product the Lie bracket is defined as usual
\be
    [F, G]_{\circ} := F\circ G - G\circ F\,.
\ee

\subsection{Higher-spin equations}

Usual HS algebra is the Lie algebra associated with associative algebra $A^0$
defined in Section \ref{HSA}.
Lie algebra $M_2$ contains usual HS algebra as a subalgebra since,
 as  is easy to see from (\ref{circ_a}),
\bee
    [F(Y), G(Y)]_{\circ} = [F(Y), G(Y)]_{\star}\,.
\eee
(To simplify notations we will use the same letter for an associative algebra and
associated Lie algebra.)
 Let us write the unfolded equations (for more detail on the unfolded formulation see,
 e.g., \cite{Vasiliev:1999}) for the  Lie algebra $M_2$
  to see the difference with the usual HS theory.

We will work with the system in  $AdS$ background described in the two-component spinor
notations  by the connection
\be
    \omega_0 = -\frac{i}{4} (\omega_0(x)^{\ga\beta} y_\ga y_\beta +
    \overline{\omega}_0(x) ^{\dot\alpha\dot{\beta}}\overline{y}_{\dot{\alpha}}
    \overline{y}_{\dot{\beta}} + 2\lambda h_0^{\alpha\dot{\beta}}(x)y_\alpha
    \overline{y}_{\dot{\beta}})
\ee
obeying the $sp(4)$ flatness conditions
\be
\dr\omega_0 + \omega_0 \star \omega_0 = 0 \,,\qquad \dr:= dx^\un \frac{\partial}{\partial x^\un}\,.
\ee
The latter can be solved, for instance, in Poincar\'e coordinates
\be
    h_{0\underline{n}}^{\alpha\dot{\beta}}(x) = z^{-1}\sigma_{\underline{n}}^{\alpha\dot{\beta}}\,,\qquad
    \omega_{0\underline{n}}^{\alpha\alpha} = -\lambda^2 z^{-1}\sigma_{\underline{n}}^{\alpha\dot{\beta}}x^\alpha_{\dot{\beta}}\,
\ee
where we use notation assuming symmetrization over upper indices denoted
by the same letter $\ga$.

The dynamical fields we are interested in  are described by the zero-form
\be
    C_{hs} = (C_{hs}(\Y|x),  C_{hs}(\Y^1; \Y^2|x))
\ee
and one-form
\be
    \omega_{hs} = (\omega_{hs}(\Y|x),  \omega_{hs}(\Y^1; \Y^2|x))\,
\ee
where the rank-two fields $\omega_{hs}(\Y^1; \Y^2|x)$ and
$C_{hs}(\Y^1; \Y^2|x)$ obey  conditions (\ref{1hs}) and (\ref{0hs})
with respect to each argument
\be
\go_{hs}(Y_1;K_1; Y_2; K_2 |x)=\go_{hs}(Y_1;-K_1; Y_2; K_2| x)
=\go_{hs}(Y_1;K_1; Y_2; -K_2| x)\,,
\ee
\be
\label{C2tw}
C_{hs}(Y_1;K_1; Y_2; K_2 |x)=-C_{hs}(Y_1;-K_1; Y_2; K_2| x)= -
C_{hs}(Y_1;K_1; Y_2; -K_2| x)\,.
\ee

Let us start with the equations on zero-forms in the twisted adjoint HS module
setting
\be
C_{hs}(Y;K|x) := \sum_{p+q=1}C_{pq}(Y|x) k^p \bar k^q\,,
\ee
\be
C_{hs}(Y^1; Y^2;K^1; K^2|x) := \sum_{\substack{p_1+q_1=1\\ p_2+q_2=1}}C_{p_1 p_2 q_1 q_2}(Y^1; Y^2|x) k^{1p_1} k^{2p_2}\bar k^{1q_1} \bar k^{2q_2}\,.
\ee
The field equations on the zero-forms $C_{hs}$ read as
\be
\label{twis}
    \dr C_{hs} + \omega_0 \circ C_{hs} - C_{hs} \circ
    \omega_0 = 0\,,
\ee
where
\be
    C_{hs} = C_{hs}(Y;K|x) + C_{hs}(Y^1; Y^2;K^1; K^2|x)\,.
\ee

From (\ref{twis}) we  obtain that the equation in the $p+q=1$ sector of $C_{hs}(Y;K|x)$ is
\be
\label{twist_deriv}
    D_{tw}C_{pq}(Y|x) = 0\,,
\ee
where
\be
    D_{tw}C_{pq}(Y|x) := D^{L}C_{pq}(Y|x) - i\lambda h_0^{\alpha\dot{\beta}}(y_{\alpha}\overline{y}_{\dot{\beta}} - \partial_{\alpha}\overline{\partial}_{\dot{\beta}})C_{pq}(Y|x)\,,
\ee
\be
    D^{L}C_{pq}(Y|x) := \dr C_{pq}(Y|x) + (\omega_0^{\alpha\beta}y_\alpha\partial_\beta + \overline{\omega}_0^{\dot{\alpha}\dot{\beta}}\overline{y}_{\dot{\alpha}}\overline{\partial}_{\dot{\beta}} )C_{pq}(Y|x)\,.
\ee
Equation (\ref{twist_deriv}) is the usual equation for massless fields in the twisted adjoint representation obtained originally in \cite{Vasiliev:1988sa} (see also \cite{Vasiliev:1999}).

In sector $p_1 + q_1 = 1$, $p_2 + q_2 = 1$ we get
\be
\label{twis_current}
    (D^{L} - i\lambda h_0^{\alpha\dot{\beta}}(y^1_{\alpha}\overline{y}^1_{\dot{\beta}}
      + y^2_{\alpha}\overline{y}^2_{\dot{\beta}} -
      \partial^1_{\alpha}\overline{\partial}^1_{\dot{\beta}} -
      \partial^2_{\alpha}\overline{\partial}^2_{\dot{\beta}}))
      C_{p_1p_2q_1q_2}(y^1, \overline{y}^1; y^2, \overline{y}^2|x) = 0\,.
\ee
This equation exactly matches the rank-two equation on currents of  \cite{Gelfond:2003vh}. Thus the part
of zero-forms with two arguments can be interpreted as describing symmetrized
conserved currents.

Since field equations (\ref{twist_deriv}) and (\ref{twis_current}) are insensitive
to the choice of $p$, $p_1$, $p_2$, $q$, $q_1$, $q_2$ obeying $p+q=1$, $p_1+q_1=1$, $p_2+q_2=1$ in the sequel   these labels will be discarded.

The current deformation of the HS equations found in \cite{Gelfond:Current}, suggests
the deformed form  of HS equations
relating zero-forms to one-forms
\be
    \begin{cases}
    \label{mult_sys}
        D_{ad}\omega(Y) = L(C(Y)) + Q(C(Y), \omega(Y)) + \Gamma^{loc}_{\eta\overline{\eta}}(C(Y^1; Y^2);Y)\\
        D_{tw}C(Y) + [\omega, C(Y)]_\star = -\mathcal{H}^{loc}_{\eta cur}(C(Y^1; Y^2);Y) - \mathcal{H}^{loc}_{\overline{\eta} cur}(C(Y^1; Y^2);Y) \,,
    \end{cases}
\ee
where
\be
    D_{ad}\omega = D^L \omega + \lambda h_0^{\alpha\dot{\beta}}
    (y_\alpha\overline{\partial}_{\dot{\beta}} + \partial_{\alpha}\overline{y}_{\dot{\beta}}) \omega\,,
\ee
\be
    L(C(Y)) = \frac{i}{4}\left(\eta \overline{H}^{\dot{\alpha}\dot{\beta}}\overline{\partial}_{\dot{\alpha}}\overline{\partial}_{\dot{\beta}}C(0, \overline{y}|x) + \overline{\eta} H^{\alpha\beta}\partial_\alpha\partial_\beta C(y, 0|x)\right)\,,
\ee
\bee
        Q(C, \omega) = \eta \int dSdT \exp(iS_AT^A)\int_0^1d\tau\Big( h(t, \tau\overline{t} - \overline{s})\omega((1 - \tau)y + s, \overline{y} + \overline{s})C(\tau t, \overline{y} + \overline{t}|x) + \nn\\
        + h(s, \overline{s}\tau - \overline{t})C(-\tau s, \overline{y} + \overline{s}|x)\omega(-(1-\tau)y - t, \overline{y} + \overline{t})\Big) + c.c.\,,\quad,
\eee

\bee
        \Gamma^{loc}_{\eta\overline{\eta}}(C(Y^1; Y^2);Y) = \frac{i}{8}\eta\overline{\eta}\int\frac{d^4\tau}{\tau_4^2}\delta(1 - \tau_3 - \tau_4)\delta'(1 - \tau_1 - \tau_2)\theta(\tau_1)\theta(\tau_2)\theta(\tau_3)\theta(\tau_4) \nn \\
        \Big(\overline{H}^{\dot{\alpha}\dot{\beta}}\overline{\partial}_{\dot{\alpha}}\overline{\partial}_{\dot{\beta}}\exp(i\tau_3\overline{\partial}_{1\dot{\alpha}}\overline{\partial}_2^{\dot{\,\alpha}} )C(\tau_1 y, \tau_4 \tau_2 \overline{y};  -\tau_2 y, -\tau_4 \tau_1 \overline{y}) + \nn \\
        +H^{\alpha\beta}\partial_{\alpha}\partial_{\beta}\exp (i\tau_3\partial_{1\alpha}\partial_2^\alpha)C(\tau_4 \tau_1 y , \tau_2 \overline{y};-\tau_4 \tau_2 y, -\tau_1\overline{y})\Big)\,,
\eee
\be
    \mathcal{H}^{loc}_{\eta cur}(C(Y^1; Y^2);Y) = \frac{1}{2}\eta \exp(i\overline{\partial}_{1\dot{\beta}}\overline{\partial}_2^{\dot{\beta}})\int_0^1 d\tau h(y, (1 - \tau)\overline{\partial}_1 - \tau \overline{\partial}_2)C(\tau y,  \overline{y}; -(1 - \tau)y, \overline{y})\,,
\ee
\be
    \mathcal{H}^{loc}_{\overline{\eta} cur}(C(Y^1; Y^2);Y) = \frac{1}{2}\overline{\eta} \exp(i\partial_{1\beta}\partial_2^{\beta})\int_0^1 d\tau h((1 - \tau)\partial_1 - \tau \partial_2, \overline{y})C(y, \tau\overline{y}; y, - (1 - \tau)\overline{y})
\ee
where $\eta$  is a complex deformation parameter, $\p_{i\ga}$ and $\bar \p_{i\dga}$ are
 derivatives over the $i$th undotted and dotted  spinor arguments,
   respectively, with upper indices and
\be
    h(a, \overline{b}) := h_0^{\alpha\dot{\beta}}a_\alpha \overline{b}_{\dot{\beta}}\,, \qquad H^{\ga\gb} := h_0^{\ga}{}^\pa  h_0^\gb{}_\pa\,,\qquad
\overline {H}^{\pa\pb} := h_0^{\ga}{}^\pa h_{0\ga}{}^{\pb}\,.
\ee

The difference between the system (\ref{mult_sys}) and that of \cite{Gelfond:Current} is that
in our case the rank-two fields $C(Y^1;Y^2)$ are algebraically independent from
 the rank-one fields $C(Y)$ while in \cite{Gelfond:Current} the former were realized
 as bilinear combinations of the latter. Hence, in \cite{Gelfond:Current} the rank-two
 fields were indeed describing the conformal currents built from the dynamical rank-one fields $C(Y)$.
 In the system described in this section the fields $C(Y^1;Y^2)$ that appear exactly on
the same place as usual currents are independent fields that are expressed via derivatives
of $C(Y)$ by virtue of equations (\ref{mult_sys}). In fact, this means that the resulting system
can be interpreted as a conformal off-shell deformation of the usual massless equations. As such it is anticipated to be related to the off-shell HS equations proposed recently  in \cite{Misuna:2019ijn}.

Indeed, for instance in the spin-one sector the deformation of Maxwell equations is
\be
\label{FJ}
\p_\un F^{\un\um} = 4\pi J^{\um}\q F = \dr A\,.
\ee
In the system considered in this paper $J^\um$ is a component of the rank-two field
 $C(Y^1;Y^2)$ independent of the rank-one fields $C(Y)$ where the two-form field
 strength $F$ lives as well as all other massless fields of the system. Hence,
equation (\ref{FJ}) simply expresses the rank-two field $J^\um$ via the rank-one field $F(Y)$.
 In \cite{Gelfond:Current} the currents like $J^\um$
were not independent, being expressed via bilinears of the rank-one fields $C(Y)$ thus describing
a nonlinear deformation of the equations of motion for massless fields.

Note that formal consistency of Maxwell equations demands the current $J$ be conserved
\be
\partial_n J^n = 0\,.
\ee
The situation with higher spins is analogous. The respective HS conservation conditions
are the only differential conditions obeyed by the primary current fields obeying
the rank-two equations \cite{Gelfond:2003vh}.
More precisely, the deformation (\ref{mult_sys}) is conformal invariant
which means that the physical HS currents are traceless while the modules of the
Poincare algebra considered in \cite{Misuna:2019ijn} include the traceful components of the
conserved currents as well. As shown in \cite{Misuna:2019ijn}, the traceful components of the
currents do not contribute to the equations for the traceless ones. This means that
the system considered in this paper remains off-shell in the conformal zero-form sector. It would be interesting to
reach more precise interpretation of the results of   \cite{Misuna:2019ijn} in terms of
conformal symmetry.

Group theoretically equations (\ref{mult_sys}) describe a nontrivial deformation of the direct sum of the
massless modules realized by the rank-one fields $C(Y)$ and the rank-two field
 $C(Y^1;Y^2)$ with the deformation parameter $\eta$. At $\eta=0$ the system decomposes into
 two independent subsystems of rank-one and rank-two fields while at $\eta\neq 0$ the system
describes the semi-direct indecomposable sum of the two models, becoming off-shell.

 Let us note that the full set of field variables also includes fields
 $\omega(Y^1; Y^2)$. The undeformed  equation for the one-forms starts with the
 flatness condition
 \be
    \dr \omega(Y^1; Y^2) + \omega(Y^1; Y^2) \circ \wedge \omega(Y^1; Y^2)  = 0\,.
 \ee
Its linearized version resulting from  (\ref{circ_a})-(\ref{circ_b}) reads as
 \be
    D^L\omega(y^1, \overline{y}^1; y^2, \overline{y}^2) +
    \lambda h_0^{\alpha\dot{\beta}}(y^1_\alpha\overline{\partial}^1_{\dot{\beta}} +
    \partial^1_\alpha\overline{y}^1_{\dot{\beta}} + y^2_\alpha\overline{\partial}^2_{\dot{\beta}} +
    \partial^2_\alpha\overline{y}^2_{\dot{\beta}})\omega(y^1, \overline{y}^1; y^2, \overline{y}^2) = 0\,.
 \ee
We anticipate that the nontrivial deformation of these equations including equations (\ref{mult_sys})
will result from the nonlinear equations of \cite{Vasiliev:2018zer} very much as
the nonlinear deformation of usual HS equations was derived in \cite{Gelfond:Current}
from the nonlinear HS equations of \cite{Vasiliev:1992av}.

\section{$M^2(A)$ system}
\label{M2u}

    In this section we analyse the dynamical system associated with the algebra $M^2(A)$.
    As explained in Section \ref{M},
 algebra $M^2(A)$ acts on fields depending on two spinor variables $F(Y^1; Y^2)$.
 This raises the question whether it is possible to identify some of these fields with
 the rank-one fields describing massless fields
  such as background $AdS$ connection $\omega_0$.

    First we observe that Lie algebra $M^2(A)$ contains a subalgebra of polynomials
    of the form $F(Y^1; Y^2) = f(Y^1) + f(Y^2)$. Indeed,
    \bee
        [f(Y^1) + f(Y^2), g(Y^1) + g(Y^2)]_\circ = \nn \\ =f(Y^1)\star_{11}g(Y^1) + f(Y^1)g(Y^2) + f(Y^2)g(Y^1) + f(Y^2)\star_{22}g(Y^2) -  \nn \\
        - g(Y^1)\star_{11}f(Y^1) - g(Y^1)f(Y^2) - g(Y^2)f(Y^1) - g(Y^2)\star_{22}f(Y^2) =  \nn \\
        =h(Y^1) + h(Y^2) \,,\quad
    \eee
    \be
        h(Y) = [f(Y), g(Y)]_\star\,.
    \ee

    This suggests to try to identify the rank-one  fields as
    \be
        \omega_0(Y^1; Y^2) = \omega_0(Y^1) + \omega_0(Y^2)\,.
    \ee
    However, this does not work in the zero-form sector. Indeed, let us decompose
     the zero-form $C$ into two parts: one containing only polynomials of $Y^1$ or $Y^2$ separately
     and the other one containing  elements that depend nontrivially on both $Y^1$ and $Y^2$,
    \be
        C(Y^1; Y^2) = C^1(Y^1) + C^1(Y^2) + C^2(Y^1; Y^2)\q C^2(Y^1; 0)=C^2(0; Y^2) =0\,,
    \ee
    \ie
    \be
        C^1(Y^{i}) = \frac{1}{2i}\sum_{m, n\geq 0}\frac{1}{m!n!}C^1_{\alpha_1\ldots\alpha_n,\dot{\beta}_1\ldots\dot{\beta}_m}y^{i\alpha_1}\cdot_{\dots}\cdot y^{i\alpha_n}\overline{y}^{i\dot{\beta}_1}\cdot_{\dots}\cdot\overline{y}^{i\dot{\beta}_m}\,,\quad i=1, 2\,,
    \ee
    \be
        C^2(Y^1; Y^2) = \frac{1}{2i}\sum_{\substack{m+n>0\\ l+p>0}} \frac{1}{m!n!l!p!}C^2_{\alpha_1\dots\alpha_n,\dot{\beta}_1\dots\dot{\beta}_m,\gamma_1\dots\gamma_l,\dot{\delta}_1\dots\dot{\delta}_p}y^{1\alpha_1}\cdot_{\dots}\cdot y^{1\alpha_n}\overline{y}^{1\dot{\beta}_1}\cdot_{\dots}\cdot\overline{y}^{1\dot{\beta}_m}y^{2\gamma_1}\cdot_{\dots}\cdot y^{2\gamma_l}\overline{y}^{2\dot{\delta}_1}\cdot_{\dots}\cdot\overline{y}^{2\dot{\delta}_p}\,.
    \ee

    Then the  equation in the  zero-form sector is
    \be
        \dr C + \omega_0 \circ C - C\circ \tilde{\omega}_0 = 0\,,
    \ee
    where
    \be
    \label{twist}
        \tilde{f}(y, \overline{y}) := f(y, -\overline{y})\,.
    \ee
    This gives three different equations on  $C^1(Y^1)$, $C^1(Y^2)$ and $C^2(Y^1; Y^2)$:
    \be
        0 = D_{tw}^1 C^1(Y^1) - \frac{i\lambda  h_0^{\alpha\dot{\beta}}}{m!n!}\sum_{m+n>0}C^2_{\alpha_1\dots\alpha_n,\dot{\beta}_1\dots\dot{\beta}_m,\alpha,\dot{\beta}}y^{1\alpha_1}\cdot_{\dots} \cdot y^{1\alpha_n}\overline{y}^{1\dot{\beta}_1}\cdot_{\dots}\cdot \overline y^{1\dot{\beta}_m}\,,
    \ee
    \be
        0 = D_{tw}^2 C^1(Y^2) - \frac{i\lambda h_0^{\alpha\dot{\beta}}}{m!n!}\sum_{m+n>0}C^2_{\alpha_1\dots\alpha_n,\dot{\beta}_1\dots\dot{\beta}_m,\alpha,\dot{\beta}}y^{2\alpha_1}\cdot_{\dots} \cdot y^{2\alpha_n}\overline{y}^{2\dot{\beta}_1}\cdot_{\dots}\cdot \overline y^{2\dot{\beta}_m}\,,
    \ee
    \bee
        0 = (D^{L} - i\lambda h_0^{\alpha\dot{\beta}}(y^1_{\alpha}\overline{y}^1_{\dot{\beta}} + y^2_{\alpha}\overline{y}^2_{\dot{\beta}} - \partial^1_{\alpha}\overline{\partial}^1_{\dot{\beta}} -  \partial^2_{\alpha}\overline{\partial}^2_{\dot{\beta}}))C^2(y^1, \overline{y}^1; y^2, \overline{y}^2) -\nn  \\
        - i\lambda h_0^{\alpha\dot{\beta}}\left(y^1_\alpha \overline{y}^1_{\dot{\beta}}C^1(Y^2) + y^2_\alpha\overline{y}^2_{\dot{\beta}}C^1(Y^1)\right)\,,\quad
    \eee
    where
    \begin{flalign}
        &C^2(y^1, \overline{y}^1; y^2, \overline{y}^2) = \nn  \\
        & = \left(\sum_{\substack{(m, n)\neq(1, 1)\\ (l,p)\neq(1, 1)\\ m+n>0 \\l+p>0}}  \frac{1}{m!n!l!p!}C^2_{\alpha_1\dots\alpha_n,\dot{\beta}_1\dots\dot{\beta}_m,\gamma_1\dots\gamma_l,\dot{\delta}_1\dots\dot{\delta}_p}y^{1\alpha_1}\cdot_{\dots}\cdot y^{1\alpha_n}\overline{y}^{1\dot{\beta}_1}\cdot_{\dots}\cdot\overline{y}^{1\dot{\beta}_m}y^{2\gamma_1}\cdot_{\dots}\cdot y^{2\gamma_l}\overline{y}^{2\dot{\delta}_1}\cdot_{\dots}\cdot\overline{y}^{2\dot{\delta}_p}\right)\,.\quad
    \end{flalign}

    The first and second equations look analogously to the equations on zero-forms with
    nontrivial right  hand side. These  equations are formally
    consistent as follows  from their derivation and can also be checked directly.
    Naively, one might think that they also describe an off-shell version of the massless  system.
    This is however not the case because of the last term in the third equation which means that
    the fields $C^1$ source  $C^2$. In turn, this means that the rank-two fields $C^2 (Y^1; Y^2)$ form
    an irreducible module that admits no limiting procedure allowing to put the massless system on
    the mass shell. This fact is in agreement with the analysis of \cite{Vasiliev:2012tv} where it was
    explained that rank-two fields carry lowest energies inappropriate for the description of massless
    fields. The field $C^2(Y^1; Y^2)$ describes the module of conserved currents which are not linked to
    the  massless fields as a result of factorization of the ideal in $M_2$. Note that,
    as mentioned in Section \ref{M}, $M^2$ is isomorphic to the Weyl algebra with the doubled set of
    generators (oscillators).

    \section{Conclusion}
  \label{con}
        In this paper we have derived the product law for the simplest multi-particle HS algebra $M_2(A)$
        and its factor-algebra $M^2(A)$ with the rank-one fields factored out. We analysed the equations of the
         system resulting from these algebras.
        We conclude that the system for $M_2(A)$ describes the off-shell HS systems with the rank-two
        fields describing conformal off-shell degrees of freedom of the originally massless system. On the other hand,
        the system resulting from $M^2(A)$ describes the infinite system of conserved currents unrelated to
        massless fields. Note that the off-shell completion of massless field equations was recently
        proposed in \cite{Misuna:2019ijn}. It would be interesting to compare the construction of \cite{Misuna:2019ijn} with the one found in this paper and, especially, to clarify
        the role of the traceful components of the currents considered in   \cite{Misuna:2019ijn}.

Analogously, one can consider multi-particle algebras of higher ranks $k$. We expect that
unfactorised algebras also describe off-shell HS systems.  One of the features illustrated by the
analysis of this paper is that
all multi-particle algebras of finite ranks have the property that higher-rank fields do not contribute
to the field equations of lower-rank fields. However, as shown in the same papers, this  is not
 true for the full  multi-particle algebra of infinite rank. This property is crucial for the
 spontaneous breaking of HS symmetries.

\section*{Acknowledgments}
 ID is particularly grateful to Olga Gelfond and Anatoly Korybut for clarification of some
  questions  in the process of calculations. MV is grateful to Nikita Misuna for the
 useful  discussion of his results on the off-shell formulation of massless fields.
   This work was supported by
the Russian Science Foundation grant 18-12-00507.

    \section*{Appendix. Derivation of $M_2(A)$ higher-spin product law}
    Let us explicitly  write down the realization of the elements of the algebra $M_2(A)$
    in terms of elements of the HS algebra $A$.

    The basis of $M_2(A)$  consists of the elements
    \be
        1, \quad \alpha_i, \quad\alpha_i\cdot\alpha_j = \alpha_j\cdot\alpha_i\,, \quad i,j=1,\dots, n\,.
    \ee
    Here  the dot product  replaces indices $1$ and $2$ of the $y$-arguments and is introduced to distinguish it from the product of elements of the original HS star-product algebra realized  by
    the polynomials of one spinor argument $F(Y)$.

     Other way around the dot product can be replaced by rewriting  formulas using two arguments $y^1$ and $y^2$. For instance, for the element $\alpha_i\cdot \alpha_j$ with $\alpha_i=F(y)$ and $\alpha_j=G(y)$
    \be
    \label{equivalence_relation}
        \alpha_i\cdot\alpha_j \sim F(y^1) G(y^2) + F(y^2) G(y^1)\,.
    \ee
    Then the product law can be checked easily for any monomial
    \bee
    \label{dot_equation}
        y_{i_1}\dots y_{i_n} \cdot y_{j_1}\dots y_{j_m}\circ y_{k_1}\dots y_{k_p}\cdot y_{l_1}\dots y_{l_o} = \nn \\ (y_{i_1}\dots y_{i_n} \star y_{k_1}\dots y_{k_p})\cdot(y_{j_1}\dots y_{j_m} \star y_{l_1}\dots y_{l_o}) + (y_{i_1}\dots y_{i_n} \star y_{l_1}\dots y_{l_o}) \cdot (y_{j_1}\dots y_{j_m} \star y_{k_1}\dots y_{k_p}) \,.\quad
    \eee
    The same can be done in terms of $y^1$ and $y^2$ using (\ref{circ_b})
    \bee
    &&  \!\!\!\!\! \!\!\!\!\! \!\!\!\!\!\Bigg((y^1_{i_1}\dots y^1_{i_n})(y^2_{j_1}\dots y^2_{j_m}) +
        (y^2_{i_1}\dots y^2_{i_n})(y^1_{j_1}\dots y^1_{j_m})\Bigg) \circ
        \Bigg((y^1_{k_1}\dots y^1_{k_p})( y^2_{l_1}\dots y^2_{l_o}) +
        (y^2_{k_1}\dots y^2_{k_p})( y^1_{l_1}\dots y^1_{l_o})\Bigg) = \nn \\
        &&\qquad= (y^1_{i_1}\dots y^1_{i_n} \star_{11} y^1_{k_1}\dots y^1_{k_p})
        (y^2_{j_1}\dots y^2_{j_m} \star_{22} y^2_{l_1}\dots y^2_{l_o}) + \nn \\
        &&\qquad+(y^1_{i_1}\dots y^1_{i_n} \star_{11} y^1_{l_1}\dots y^1_{l_o})
        (y^2_{j_1}\dots y^2_{j_m} \star_{22} y^2_{k_1}\dots y^2_{k_p}) + \nn \\
        &&\qquad+ (y^2_{i_1}\dots y^2_{i_n} \star_{22} y^2_{l_1}\dots y^2_{l_o})
        (y^1_{j_1}\dots y^1_{j_m} \star_{11} y^1_{k_1}\dots y^1_{k_p}) + \nn \\
       &&\qquad +(y^2_{i_1}\dots y^2_{i_n} \star_{22} y^2_{k_1}\dots y^2_{k_p})
        (y^1_{j_1}\dots y^1_{j_m} \star_{11} y^1_{l_1}\dots y^1_{l_o}).\label{index_equation} \quad
    \eee
    So the right hand sides of (\ref{dot_equation}) and (\ref{index_equation}) coincide up to equivalence  (\ref{equivalence_relation}). This proves (\ref{circ_b}).

\end{document}